\documentclass[letterpaper, 10 pt, conference]{ieeeconf}
\IEEEoverridecommandlockouts
\usepackage{amsfonts,amsmath,amssymb} 

\usepackage{psfrag,color}
\usepackage{enumerate,cite,latexsym,graphicx}
\newtheorem{theorem}{Theorem}
\newtheorem{lemma}{Lemma}

\newtheorem{definition}{Definition}
\newtheorem{observation}{Observation}

\def\tr{\mathop{\rm Tr}\nolimits} 

\title{\LARGE \bf  A Popov Stability Condition for Uncertain Linear Quantum Systems
}

\author{ Matthew R~James, Ian R.~Petersen and Valery Ugrinovskii%
\thanks{This work was supported by the
Australian Research Council (ARC) under projects DP110102322,
DP1094650 and FL110100020,  and the Air Force Office of Scientific
Research (AFOSR). This material is based on research sponsored by the
Air Force Research Laboratory, under agreement number
FA2386-09-1-4089.  The U.S. Government is authorized to reproduce and
distribute reprints for Governmental purposes notwithstanding any
copyright notation thereon.
The views and conclusions contained herein are those of the authors
and should not be interpreted as necessarily representing the official
policies or endorsements, either expressed or implied, of the Air
Force Research Laboratory or the U.S. Government. }%
\thanks{Matthew R. James is with the ARC Centre for Quantum
  Computation and Communication Technology, Research School of  Engineering,
College of Engineering and Computer Science, The Australian National University, Canberra, ACT 0200,
Australia. Email: Matthew.James@anu.edu.au.}
\thanks{Ian R. Petersen and Valery Ugrinovskii are with the School of  Engineering and Information Technology, 
        University of New South Wales at the Australian Defence Force Academy, Canberra ACT 2600, Australia.
         {\tt\small \{i.r.petersen,v.ugrinovskii\}@gmail.com}. The
         work of Valery Ugrinovskii was carried out in part while he
         was a visitor at the Australian National University.} 
}%

\begin{document}

\maketitle
\thispagestyle{empty}
\pagestyle{empty}

\begin{abstract}
This paper considers a Popov type approach to the problem of robust stability for a class of 
uncertain linear quantum systems subject to unknown perturbations in the
system Hamiltonian. A general stability result is given for
a general class of perturbations to the system Hamiltonian. Then, the
special case of a nominal linear quantum system is considered with
 quadratic perturbations to the system
Hamiltonian. In this case, a robust stability condition is given in terms of a frequency domain condition which is of the same form as the standard Popov stability condition. 
\end{abstract}

\section{Introduction} \label{sec:intro}
This paper builds on the previous papers \cite{PUJ1a,PUJ2,PUJ3a} which
consider the problem of robust stability analysis for open 
quantum systems subject to perturbations in either the system
Hamiltonian or coupling operator, which together define the dynamics
of the quantum system. The results of these papers can be regarded as
extensions of the classical small gain theorem for robust stability 
 to the case of quantum systems. The main
contribution of this paper is a result which can be regarded as an
extension of the classical Popov criterion for absolute stability to
the case of open quantum systems. In particular, we extend the result
of \cite{PUJ2}, in which the perturbations to the system Hamiltonian
are uncertain quadratic perturbations, to obtain a corresponding Popov
robust stability result. 

The small gain theorem and the Popov criterion for absolute stability
are two of the most useful  tests for  robust stability and nonlinear system
stability; e.g., see \cite{KHA02}. Both of these stability tests
consider a Lur'e system which is the feedback interconnection between
a linear time invariant system and a sector bounded nonlinearity or
uncertainty. The key distinction between the  small gain theorem and
the Popov criterion is that the small gain theorem establishes
absolute stability via the use of a fixed quadratic Lyapunov function
whereas the Popov criterion relies on a Lyapunov function of the Lur'e
Postnikov form which  involves the sum of a quadratic term and a term
dependent on the integral of the nonlinearity itself. The small gain
theorem can be used to establish stability in the presence of
time-varying uncertainties and nonlinearities whereas the Popov
criterion only applies to static time-invariant
nonlinearities. However, the Popov criterion is less conservative
than the small gain theorem. Hence, we are motivated to obtain a quantum Popov stability criterion in order to obtain less conservative results. 

The study of quantum feedback control theory has been the subject of increasing interest in recent years; e.g., see
\cite{YK03A,YK03B,YAM06,JNP1,NJP1,GGY08,MaP3,MaP4,YNJP1,GJ09,GJN10,WM10,PET10Ba}. In
particular, the papers \cite{GJ09,JG10} consider a framework of
quantum systems defined in terms of a triple $(S,L,H)$ where $S$ is a
scattering matrix, $L$ is a vector of coupling operators and $H$ is a
Hamiltonian operator. The paper \cite{JG10} then introduces notions of
dissipativity and stability for this class of quantum systems.  As in the papers \cite{PUJ1a,PUJ3a}, the results of this paper build on the stability results of \cite{JG10} to obtain robust 
stability results for uncertain  quantum systems in which the quantum system Hamiltonian is
decomposed as $H =H_1+H_2$ where $H_1$ is a known nominal Hamiltonian
and $H_2$ is a perturbation Hamiltonian, which is contained in a
specified set of Hamiltonians $\mathcal{W}$. 

For this general class of uncertain quantum systems, the paper first obtains a general abstract version of the Popov stability criterion which requires finding a Lyapunov type operator to satisfy an operator inequality. The paper then considers the case in which the
nominal Hamiltonian $H_1$ is a quadratic function of annihilation and
creation operators and the coupling operator vector is a linear
function of annihilation and creation operators. This case corresponds
to a nominal  linear quantum system; e.g., see
\cite{JNP1,NJP1,MaP3,MaP4,PET10Ba}. Also, it is assumed that the
perturbation Hamiltonian is quadratic but uncertain. In this special case, a robust stability
stability criterion is obtained in terms of a frequency domain condition which takes the same form as the classical Popov stability criterion. 

The remainder of the paper proceeds as follows. In Section
\ref{sec:systems}, we define the general class of uncertain quantum
systems under consideration. In this section, we also present a general Popov type stability result for this class of quantum systems. In Section
\ref{sec:quad_pert}, we consider a  class of uncertain quadratic perturbation Hamiltonians. In Section
\ref{sec:linear}, we specialize to the case of linear nominal
quantum systems and obtain a  robust stability result for this case
in which the stability condition is a frequency domain condition in
the same form as the classical Popov stability condition. In Section
\ref{sec:example} we present an illustrative example involving a
quantum system arising from an optical parametric amplifier. In Section \ref{sec:conclusions},
we present some conclusions. 
\section{Quantum Systems} \label{sec:systems}
We consider  open quantum systems defined by  parameters $(S,L,H)$ where $H = H_1+H_2$; e.g., see \cite{GJ09,JG10}.  The corresponding generator for this quantum system is given by 
\begin{equation}
\label{generator}
\mathcal{G}(X) = -i[X,H] + \mathcal{L}(X)
\end{equation}
where $ \mathcal{L}(X) =
\frac{1}{2}L^*[X,L]+\frac{1}{2}[L^*,X]L$. Here, $[X,H] = XH-HX$
denotes the commutator between two operators and the notation $^*$
denotes the adjoint  of an operator. Also, $H_1$ is a
self-adjoint operator on the underlying Hilbert space referred to as the
nominal Hamiltonian and $H_2$ is a self-adjoint operator on the underlying
Hilbert space referred to as the perturbation Hamiltonian.  The triple
$(S,L,H)$, along with the corresponding generators define the Heisenberg
evolution $X(t)$ of an operator $X$ according to a quantum stochastic
differential equation; e.g., see \cite{JG10}. 

\begin{figure}[htbp]
\begin{center}
\includegraphics[width=6cm]{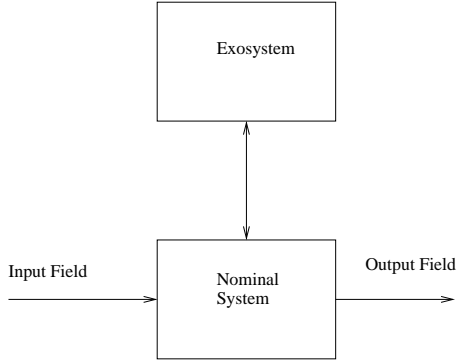}
\end{center}
\caption{Block diagram representation of an open quantum system interacting with an exosystem.}
\label{F1}
\end{figure}

The problem under consideration involves establishing robust stability
properties for an uncertain open quantum system for the case in which the perturbation Hamiltonian is contained in a given set $\mathcal{W}_1$. Using the notation of \cite{JG10}, the set $\mathcal{W}_1$ defines a set of exosystems. This situation is illustrated in the block diagram shown in Figure \ref{F1}. 
The main robust stability results presented in this paper will build on the following result from \cite{JG10}. 
\begin{lemma}[See Lemma 3.4 of \cite{JG10}.]
\label{L00}
Consider an open quantum system defined by $(S,L,H)$ and suppose there exists a non-negative self-adjoint operator $V$ on the underlying Hilbert space such that
\begin{equation}
\label{lyap_ineq0}
\mathcal{G}(V) + cV \leq \lambda
\end{equation}
where $c > 0$ and $\lambda$ are real numbers. Then for any plant state, we have
\[
\left<V(t)\right> \leq e^{-ct}\left<V\right> + \frac{\lambda}{c},~~\forall t \geq 0.
\]
Here $V(t)$ denotes the Heisenberg evolution of the operator $V$ and $\left<\cdot\right>$ denotes quantum expectation; e.g., see \cite{JG10}.
\end{lemma}
We will also use
following result, which is a slight modification of Theorem 3.1 and Lemma
3.4 of \cite{JG10}.  
\begin{lemma}
\label{L0}
Consider an open quantum system defined by $(S,L,H)$ and suppose there
exists non-negative self-adjoint operators $V$ and $W$ on the underlying
Hilbert space such that the following \emph{quantum dissipation inequality}
holds  
\begin{equation} 
\label{lyap_ineq}
\mathcal{G}(V) + W \leq \lambda
\end{equation}
where $\lambda$ is a real number. Then for any plant state, we have
\begin{equation}
\limsup_{T\to\infty}\frac{1}{T}\int_0^T \langle W(t)\rangle dt \leq
\lambda. 
\label{stability}
\end{equation}
Here $W(t)$ denotes the Heisenberg evolution of the operator $W$ and
$\langle\cdot\rangle$ denotes quantum expectation; e.g., see \cite{JG10}.  
\end{lemma}

\emph{Proof }
The proof is similar to the proof of Lemma 3.4 in \cite{JG10}. In a similar
manner we obtain from (\ref{lyap_ineq})
\begin{equation}
\mathbb{E}_t \left[V(t+h)-V(t)+\int_t^{t+h}W(s)ds\right]\le \lambda h.
\end{equation}
where $\mathbb{E}_t$ denotes vacuum expectation operator. Taking the vacuum
expectation $\mathbb{E}_0$ on both sides, and noting that
$\mathbb{E}_0\mathbb{E}_t=\mathbb{E}_0$ results in
\begin{eqnarray*}
\langle \psi,\mathbb{E}_0[V(t+h)]\psi\rangle -\langle
\psi,\mathbb{E}_0[V(t)]\psi\rangle \\
+\int_t^{t+h}\langle
\psi,\mathbb{E}_0[W(s)]\psi\rangle ds\le \lambda h \langle
\psi,\psi\rangle 
\end{eqnarray*}
for any $\psi$ in the underlying Hilbert space. Then for any corresponding
plant state (i.e., for $\psi$ such that  $\langle
\psi,\psi\rangle=1$) we have  
\begin{equation*}
\frac{d}{dt}\langle V(t) \rangle +\langle W(t)\rangle \le \lambda.
\end{equation*}
Then (\ref{stability}) follows in a standard manner. \hfill $\Box$

\subsection{Commutator Decomposition}
We now consider a set of self-adjoint perturbation Hamiltonians
$H_2 \in \mathcal{W}_1$. For a given set of non-negative self-adjoint operators 
$\mathcal{P}$, a set of Popov scaling parameters $\Theta \subset [0,\infty)$, a self-adjoint operator $H_1$, which is the nominal Hamiltonian, 
a  coupling operator $L$, and   for  a real parameter  $\beta \geq 0$,  this set $\mathcal{W}_1$ is defined in terms of the commutator decompositions
\begin{eqnarray}
\label{alt_comm_condition}
[V-\theta H_1,H_2] &=& [V-\theta H_1,z^\dagger]w
-w^\dagger[z,V-\theta H_1],\nonumber \\
\mathcal{L}(H_2)&\le &\mathcal{L}(z^\dagger)w+w^\dagger \mathcal{L}(z)
+ \beta [z,L]^\dagger [z,L] \nonumber \\  
\end{eqnarray}
for  all $V \in \mathcal{P}$ and $\theta \in \Theta$, where $w$ and $z$ are given 
operator vectors of the same dimension. Here, the notation $^\dagger$ denotes the adjoint transpose of a vector of
operators. In addition, the notation $^\#$ denotes the vector of adjoint operators for a given vector of operators. 

 Then, the set $\mathcal{W}_1$ will be defined in terms of the
sector bound condition
\begin{eqnarray}
\label{sector2a}
\left(w-\frac{1}{\gamma} z\right)^\dagger \left(w-\frac{1}{\gamma}z \right) &\leq& \frac{1}{\gamma^2}z^\dagger z
\end{eqnarray}
where $\gamma > 0$ is a given constant. That is, we define
\begin{equation}
\label{W2}
\mathcal{W}_1= \left\{\begin{array}{l}H_2 \geq 0: \exists ~w,~z, \mbox{
      such that 
(\ref{sector2a})   } \\
\mbox{  and (\ref{alt_comm_condition}) are satisfied } 
\forall V \in \mathcal{P},~\theta \in \Theta\end{array}\right\}.
\end{equation}

Using this definition, we obtain the following theorem. 
\begin{theorem}
\label{T2}
Consider a set of non-negative self-adjoint operators $\mathcal{P}$,
 an open quantum system $(S,L,H)$ and an observable $W$ where
$H=H_1+H_2$ and $H_2 \in \mathcal{W}_1$ defined in (\ref{W2}). 
Suppose there exists a $V\in
\mathcal{P}$ and real constants
 $\theta \in \Theta$,  
$\tilde \lambda \geq 0$ such that 
\begin{eqnarray}
\label{dissip1a}
\lefteqn{-i[V,H_1]+ \mathcal{L}(V)} && \nonumber \\
&&+\frac{1}{\gamma}\left(i[z,V-\theta
  H_1]+\theta\mathcal{L}(z)+z\right)^\dagger \nonumber \\
&&\times \left(i[z,V-\theta H_1]+\theta\mathcal{L}(z)+z\right)
\nonumber \\ 
&&+ \theta \beta[z,L]^\dagger [z,L]+ W \leq \tilde \lambda.
\end{eqnarray}
Then 
\begin{eqnarray*}
\limsup_{T\to\infty}\frac{1}{T}\int_0^T\langle W(t)\rangle dt \leq  \tilde \lambda.
\end{eqnarray*}
Here $W(t)$ denotes the Heisenberg evolution of the operator $W$.
\end{theorem}

\noindent
{\em Proof:}
Let $\theta \in \Theta$ and 
$\tilde \lambda \geq 0$   be given such that the conditions of the theorem are satisfied  and consider $\mathcal{G}(V+\theta H_2)$ defined in (\ref{generator}). Then 
\begin{eqnarray}
\label{ineq1a.1}
\mathcal{G}(V+\theta H_2) &=& -i[V+\theta H_2,H_1+H_2]+ \mathcal{L}(V+\theta H_2)
\nonumber \\
&=&
-i[V,H_1]-i\theta[H_2,H_1]-i[V,H_2]\nonumber \\
&&-i\theta[H_2,H_2]+\mathcal{L}(V)+\theta\mathcal{L}(H_2) \nonumber \\
&=& -i[V,H_1]-i[V-\theta H_1,H_2] + \mathcal{L}(V) \nonumber \\
&&+\theta\mathcal{L}(H_2).
\end{eqnarray}
Using the decomposition in the first equation (\ref{alt_comm_condition}), we
have  
\begin{eqnarray}
\label{ineq1a.2}
\mathcal{G}(V+\theta H_2) &=& 
-i[V,H_1]+ \mathcal{L}(V) -i[V-\theta H_1,z^\dagger]w \nonumber \\
&&+iw^\dagger[z,V-\theta H_1]+ \theta\mathcal{L}(H_2).
\end{eqnarray}
Now 
\[
[V-\theta H_1,z^\dagger]^\dagger = z(V-\theta H_1)-(V-\theta
H_1)z=[z,V-\theta H_1]
\] 
since $V-\theta H_1$ is self-adjoint. This confirms that the operator on
the right hand side of the above identity is   
a self-adjoint operator. Therefore, the following inequality follows from
the second equation (\ref{alt_comm_condition}):
\begin{eqnarray}
\label{ineq1a}
\lefteqn{\mathcal{G}(V+\theta H_2)} && \nonumber \\
&\le & -i[V,H_1]+ \mathcal{L}(V) \nonumber \\
&&-i[V-\theta H_1,z^\dagger]w +iw^\dagger[z,V-\theta H_1]\nonumber \\
&&+ \theta(\mathcal{L}(z^\dagger)w+w^\dagger\mathcal{L}(z))+
\theta\beta [z,L]^\dagger [z,L].
\end{eqnarray}
Also, note that 
\begin{eqnarray}
(\mathcal{L}(z))^\dagger
&=&\left(\frac{1}{2}L^\dagger[z,L]+\frac{1}{2}[L^\dagger,z]L\right)^\dagger 
\nonumber\\ 
&=&\frac{1}{2}[L^\dagger,z^\dagger]L+\frac{1}{2}L^\dagger[z^\dagger,L] 
\nonumber\\ 
&=& \mathcal{L}(z^\dagger).
\label{L*}
\end{eqnarray}
Furthermore, 
\begin{eqnarray*}
0 &\leq& \left(\frac{i[z,V-\theta H_1]+\theta\mathcal{L}(z)}{\sqrt{\gamma}}
-  \sqrt{\gamma}\left(w-z/\gamma\right)\right)^\dagger\\
&&\times\left(
\frac{i[z,V-\theta H_1]+\theta\mathcal{L}(z)}{\sqrt{\gamma}}
-  \sqrt{\gamma}\left(w-z/\gamma\right)
\right)\nonumber \\
&=&  \frac{1}{\gamma}\left(i[z,V-\theta
  H_1]+\theta\mathcal{L}(z)+z\right)^\dagger \\
&&\times \left(i[z,V-\theta H_1]+\theta\mathcal{L}(z)+z\right)\\
&&-\left(i[z,V-\theta H_1]+\theta\mathcal{L}(z)+z\right)^\dagger w\\
&&-w^\dagger\left(i[z,V-\theta H_1]+\theta\mathcal{L}(z)+z\right)
+\gamma w^\dagger w.
\end{eqnarray*}
This implies that
\begin{eqnarray*}
\lefteqn{\left(i[z,V-\theta H_1]+\theta\mathcal{L}(z)+z\right)^\dagger w}
&& \\
&&+w^\dagger\left(i[z,V-\theta H_1]+\theta\mathcal{L}(z)+z\right)
-\gamma w^\dagger w \\
&\leq& \frac{1}{\gamma}\left(i[z,V-\theta
  H_1]+\theta\mathcal{L}(z)+z\right)^\dagger \\
&&\times \left(i[z,V-\theta H_1]+\theta\mathcal{L}(z)+z\right).
\end{eqnarray*}
Using this inequality and (\ref{ineq1a}), we have 
\begin{eqnarray}
\label{ineq2a}
\lefteqn{\mathcal{G}(V+\theta H_2) + \frac{1}{\gamma}z^\dagger z
-\gamma \left(w-\frac{1}{\gamma} z\right)^\dagger \left(w-\frac{1}{\gamma}z
\right)} &&
 \nonumber \\
&\le&-i[V,H_1]+ \mathcal{L}(V) \nonumber \\
&&+\left(i[z,V-\theta H_1]+\theta\mathcal{L}(z)+ z\right)^\dagger w\nonumber \\
&&+w^\dagger \left(i[z,V-\theta H_1]+\theta\mathcal{L}(z)+z\right) \nonumber \\
&& - \gamma w^\dagger w+ \theta \beta [z,L]^\dagger [z,L]\nonumber \\
&\le&-i[V,H_1]+ \mathcal{L}(V) \nonumber \\
&&+\frac{1}{\gamma}\left(i[z,V-\theta
  H_1]+\theta\mathcal{L}(z)+z\right)^\dagger \nonumber \\
&&\times \left(i[z,V-\theta H_1]+\theta\mathcal{L}(z)+z\right)
\nonumber \\ 
&&+ \theta \beta[z,L]^\dagger [z,L].
\end{eqnarray}
Then it follows from (\ref{sector2a}) and (\ref{dissip1a}) that
\[
\mathcal{G}(V+\theta H_2) + W \leq \tilde \lambda.
\]
The result of the theorem then  follows from Lemma \ref{L0}.
\hfill $\Box$

\section{Quadratic Perturbations of the Hamiltonian}
\label{sec:quad_pert}
We consider a set $\mathcal{W}_2$ of quadratic perturbation
Hamiltonians of the  form
\begin{equation}
\label{H2quad}
H_2 = \frac{1}{2}\left[\begin{array}{cc}\zeta^\dagger &
      \zeta^T\end{array}\right]\Delta
\left[\begin{array}{c}\zeta \\ \zeta^\#\end{array}\right]
\end{equation}
where $\Delta \in \mathbb{C}^{2m\times 2m}$ is a Hermitian matrix of the
form
\begin{equation}
\label{Delta_form}
\Delta= \left[\begin{array}{cc}\Delta_1 & \Delta_2\\
\Delta_2^\# &     \Delta_1^\#\end{array}\right],
\end{equation}
$\Delta_1 = \Delta_1^\dagger$, $\Delta_2 = \Delta_2^T$ and $\zeta$ is
a given vector of operators. 
Here, in the case of complex matrices, the notation $^\dagger$ refers
to the complex conjugate transpose of a matrix. Also, in the case of complex matrices, the notation $^\#$ refers to 
the complex conjugate matrix. In addition, for this case we assume
that $\Theta =[0,\infty)$. 


The matrix $\Delta$
is subject to the  bounds
\begin{equation}
\label{Delta_bound}
0\leq \Delta \leq \frac{4}{\gamma}I.
\end{equation}
Then we define
\begin{equation}
\label{W3}
\mathcal{W}_2 = \left\{\begin{array}{l}H_2 \mbox{ of the form
      (\ref{H2quad}), (\ref{Delta_form}) such that 
} \\
\mbox{ condition (\ref{Delta_bound}) is satisfied}\end{array}\right\}.
\end{equation}
Using this definition, we obtain the following lemma. 
\begin{lemma}
\label{LA}
Suppose that  $[z,L]$ is a constant vector. Then, for any set of self-adjoint operators $\mathcal{P}$,
\[
\mathcal{W}_2 \subset \mathcal{W}_1.
\]
\end{lemma}

\noindent
{\em Proof:}
Given any $H_2 \in \mathcal{W}_2$, let 
$
z = \left[\begin{array}{c}\zeta \\ \zeta^\#\end{array}\right]
$
and
$
w = \frac{1}{2}\Delta z.
$
Hence, 
$
H_2 = w^\dagger z.
$
Then, for any  $V \in \mathcal{P}$ and $\theta \geq 0$,  let $\tilde V = V-\theta H_1$ and we have
\begin{eqnarray*}
[\tilde V,z^\dagger]w 
&=& \frac{1}{2}\left(\tilde Vz^\dagger -z^\dagger \tilde V\right)\Delta z \\
&=& \frac{1}{2}\tilde Vz^\dagger\Delta z-\frac{1}{2}z^\dagger \Delta \tilde Vz,
\end{eqnarray*}
since $\tilde V$ is a scalar operator and $\Delta$ is a constant matrix. 
Also, 
\begin{eqnarray*}
w^\dagger[z,\tilde V]
&=&
\frac{1}{2}z^\dagger \Delta \left(z\tilde V -\tilde Vz \right)\\
&=& \frac{1}{2}z^\dagger \Delta z\tilde V - \frac{1}{2}z^\dagger \Delta \tilde Vz.
\end{eqnarray*}
Hence, 
\begin{eqnarray*}
[\tilde V,z^\dagger]w - w^\dagger[z,\tilde V]
 &=& \frac{1}{2}\tilde Vz^\dagger\Delta z-\frac{1}{2}z^\dagger \Delta z\tilde V
\nonumber \\
&=&\frac{1}{2}[\tilde V,z^\dagger\Delta z] \nonumber \\
&=& [\tilde V,H_2].
\end{eqnarray*}
Similarly, 
\begin{eqnarray}
\label{comm1}
[L,z^\dagger]w - w^\dagger[z,L]=[L,H_2].
\end{eqnarray}

In addition, 
\begin{eqnarray}
\label{comm2}
 \lefteqn{w^T[z,L]^\#L-L[z,L]^\dagger w}\nonumber \\
&=& \frac{1}{2}z^T \Delta^T [z,L]^\#L - \frac{1}{2}L[z,L]^\dagger\Delta z \nonumber\\
&=&\frac{1}{2} [z,L]^\dagger \Delta z L - \frac{1}{2}L[z,L]^\dagger\Delta L z \nonumber\\
&=&\frac{1}{2} [z,L]^\dagger \Delta[z,L] \nonumber\\
&=& \frac{1}{2} [z,L]^\dagger \Delta[z,L]
\end{eqnarray}
and similarly
\begin{eqnarray}
\label{comm3}
L^*[z,L]^Tw^\# - w^\dagger [z,L]L^* &=& \frac{1}{2} [z,L]^\dagger \Delta[z,L].\nonumber \\
\end{eqnarray}
Now using (\ref{comm1}), (\ref{comm2}), (\ref{comm3}) and the assumption that  $[z,L]$ is a constant vector,
 it follows that
\begin{eqnarray*}
\mathcal{L}(H_2) &=& \frac{1}{2}L^*[H_2,L] + \frac{1}{2}[L^*,H_2]L \\
&=& -\frac{1}{2}L^*\left([L,z^\dagger]w - w^\dagger[z,L]\right)\\
&&-\frac{1}{2}\left(w^\dagger[L,z^\dagger]^\dagger-[z,L]^\dagger w\right)L\\
&=& \frac{1}{2}L^*[z^\dagger,L]w+\frac{1}{2}L^*w^\dagger[z,L]\\
&&+\frac{1}{2}w^\dagger[z^\dagger,L]^\dagger L + \frac{1}{2}[z,L]^\dagger w L\\
&=& \frac{1}{2}L^*[z^\dagger,L]w+\frac{1}{2}L^*[z,L]^Tw^\#\\
&&+\frac{1}{2}w^\dagger[z^\dagger,L]^\dagger L + \frac{1}{2}w^T[z,L]^\# L\\
&=& \frac{1}{2}L^*[z^\dagger,L]w+\frac{1}{2}w^\dagger [z,L]L^*\\
&&+\frac{1}{2}w^\dagger[z^\dagger,L]^\dagger L +\frac{1}{2}L[z,L]^\dagger w\\
&&+[z,L]^\dagger \Delta[z,L]\\
&=& \frac{1}{2}L^*[z^\dagger,L]w+\frac{1}{2}[L^*,z^\dagger] L w\\
&&+\frac{1}{2}w^\dagger L^*[z,L]+\frac{1}{2}w^\dagger[L^*,z] L \\
&&+[z,L]^\dagger \Delta[z,L]\\
&=& \mathcal{L}(z^\dagger)w+w^\dagger \mathcal{L}(z)+[z,L]^\dagger \Delta[z,L].
\end{eqnarray*}
It then follows from  (\ref{Delta_bound}) that
\[
\mathcal{L}(H_2) \leq \mathcal{L}(z^\dagger)w+w^\dagger \mathcal{L}(z)+\frac{4}{\gamma}[z,L]^\dagger[z,L].
\]
Therefore we can conclude that both of the conditions in (\ref{alt_comm_condition}) are satisfied with $\beta = \frac{4}{\gamma}$. Also,  condition (\ref{Delta_bound}) implies
\[
H_2 = w^\dagger z = \frac{1}{2}z^\dagger \Delta z \geq 0,
\]
and
\begin{eqnarray*}
\left(w-\frac{1}{\gamma} z\right)^\dagger \left(w-\frac{1}{\gamma}z \right) 
&=& w^\dagger w - \frac{1}{\gamma} z^\dagger w \\
&&-\frac{1}{\gamma}w^\dagger z + \frac{1}{\gamma^2} z^\dagger z\\
&=&\frac{1}{4}z^\dagger \Delta \Delta z - \frac{1}{\gamma} z^\dagger \Delta z\\
&&+ \frac{1}{\gamma^2} z^\dagger z\\
&\leq&\frac{1}{\gamma^2} z^\dagger z
\end{eqnarray*}
which implies (\ref{sector2a}). 
Hence, $H_2 \in
\mathcal{W}_1$. Therefore, $\mathcal{W}_2 \subset \mathcal{W}_1$.
\hfill $\Box$
\section{The Linear  Case}
\label{sec:linear}
We now consider the  case in which the nominal quantum system corresponds to a linear quantum system; e.g., see \cite{JNP1,NJP1,MaP3,MaP4,PET10Ba}. In this case, we assume that $H_1$ is of the form 
\begin{equation}
\label{H1}
H_1 = \frac{1}{2}\left[\begin{array}{cc}a^\dagger &
      a^T\end{array}\right]M
\left[\begin{array}{c}a \\ a^\#\end{array}\right]
\end{equation}
where $M \in \mathbb{C}^{2n\times 2n}$ is a Hermitian matrix of the
form
\[
M= \left[\begin{array}{cc}M_1 & M_2\\
M_2^\# &     M_1^\#\end{array}\right]
\]
and $M_1 = M_1^\dagger$, $M_2 = M_2^T$.  Here $a$ is a vector of annihilation
operators on the underlying Hilbert space and $a^\#$ is the
corresponding vector of creation operators. 
The annihilation and creation operators are assumed to satisfy the
canonical commutation relations:
\begin{eqnarray}
\label{CCR2}
\left[\left[\begin{array}{l}
      a\\a^\#\end{array}\right],\left[\begin{array}{l}
      a\\a^\#\end{array}\right]^\dagger\right]
&=&\left[\begin{array}{l} a\\a^\#\end{array}\right]
\left[\begin{array}{l} a\\a^\#\end{array}\right]^\dagger
\nonumber \\
&&- \left(\left[\begin{array}{l} a\\a^\#\end{array}\right]^\#
\left[\begin{array}{l} a\\a^\#\end{array}\right]^T\right)^T\nonumber \\
&=& J
\end{eqnarray}
where $J = \left[\begin{array}{cc}I & 0\\
0 & -I\end{array}\right]$; e.g., see \cite{GGY08,GJN10,PET10Ba}.

In addition, we assume $L$ is of the form 
\begin{equation}
\label{L}
L = \left[\begin{array}{cc}N_1 & N_2 \end{array}\right]
\left[\begin{array}{c}a \\ a^\#\end{array}\right] = \tilde N\left[\begin{array}{c}a \\ a^\#\end{array}\right]
\end{equation}
where $N_1 \in \mathbb{C}^{1\times n}$ and $N_2 \in
\mathbb{C}^{1\times n}$. Also, we write
\[
\left[\begin{array}{c}L \\ L^\#\end{array}\right] = N
\left[\begin{array}{c}a \\ a^\#\end{array}\right] =
\left[\begin{array}{cc}N_1 & N_2\\
N_2^\# &     N_1^\#\end{array}\right]
\left[\begin{array}{c}a \\ a^\#\end{array}\right].
\]

In addition, we assume that $V$ is of the form 
\begin{equation}
\label{quadV}
V = \left[\begin{array}{cc}a^\dagger &
      a^T\end{array}\right]P
\left[\begin{array}{c}a \\ a^\#\end{array}\right]
\end{equation}
where $P \in \mathbb{C}^{2n\times 2n}$ is a positive-definite Hermitian matrix of the
form
\begin{equation}
\label{Pform}
P= \left[\begin{array}{cc}P_1 & P_2\\
P_2^\# &     P_1^\#\end{array}\right].
\end{equation}
 Hence, we consider the set of  non-negative self-adjoint operators
$\mathcal{P}_1$ defined as
\begin{equation}
\label{P1}
\mathcal{P}_1 = \left\{\begin{array}{l}V \mbox{ of the form
      (\ref{quadV}) such that $P > 0$ is a 
} \\
\mbox{  Hermitian matrix of the form (\ref{Pform})}\end{array}\right\}.
\end{equation}


In the linear case, we also let $\zeta = E_1a+E_2 a^\#$ and hence we can write
\begin{equation}
\label{z01}
z = \left[\begin{array}{c}\zeta \\ \zeta^\#\end{array}\right]= 
\left[\begin{array}{cc}
E_1 & E_2\\
E_2^\# & E_1^\#
\end{array}\right]\left[\begin{array}{c}a \\ a^\#\end{array}\right] 
=E\left[\begin{array}{c}a \\ a^\#\end{array}\right].
\end{equation}

We will also consider a specific notion of robust mean square stability. 
\begin{definition}
\label{D1}
An uncertain open quantum system defined by  $(S,L,H)$ where $H=H_1+H_2$ with $H_1$ of the form (\ref{H1}), $H_2 \in \mathcal{W}$, and $L$  of the form (\ref{L}) is said to be {\em robustly mean square stable} if for any $H_2 \in \mathcal{W}$, there exist constants $c_1 > 0$, $c_2 > 0$ and $c_3 \geq 0$ such that
\begin{eqnarray}
\label{ms_stable0}
\lefteqn{\left< \left[\begin{array}{c}a(t) \\ a^\#(t)\end{array}\right]^\dagger \left[\begin{array}{c}a(t) \\ a^\#(t)\end{array}\right] \right>}\nonumber \\
&\leq& c_1e^{-c_2t}\left< \left[\begin{array}{c}a \\ a^\#\end{array}\right]^\dagger \left[\begin{array}{c}a \\ a^\#\end{array}\right] \right>
+ c_3~~\forall t \geq 0.
\end{eqnarray}
Here $\left[\begin{array}{c}a(t) \\ a^\#(t)\end{array}\right]$ denotes the Heisenberg evolution of the vector of operators $\left[\begin{array}{c}a \\ a^\#\end{array}\right]$; e.g., see \cite{JG10}.
\end{definition}

In order to address the issue of robust mean square stability for the
uncertain linear quantum systems under consideration, we first require some algebraic identities.
\begin{lemma}
\label{L2}
Given $V \in \mathcal{P}_1$, $H_1$ defined as in (\ref{H1}) and $L$ defined as in (\ref{L}), then
\begin{eqnarray*}
\lefteqn{[V,H_1] =}\nonumber \\
&& \left[\left[\begin{array}{cc}a^\dagger &
      a^T\end{array}\right]P
\left[\begin{array}{c}a \\ a^\#\end{array}\right],\frac{1}{2}\left[\begin{array}{cc}a^\dagger &
      a^T\end{array}\right]M
\left[\begin{array}{c}a \\ a^\#\end{array}\right]\right] \nonumber \\
&=& \left[\begin{array}{c}a \\ a^\#\end{array}\right]^\dagger 
\left[
PJM - MJP 
\right] \left[\begin{array}{c}a \\ a^\#\end{array}\right].
\end{eqnarray*}
Also,
\begin{eqnarray*}
\lefteqn{\mathcal{L}(V) =} \nonumber \\
&& \frac{1}{2}L^\dagger[V,L]+\frac{1}{2}[L^\dagger,V]L \nonumber \\
&=& \tr\left(PJN^\dagger\left[\begin{array}{cc}I & 0 \\ 0 & 0 \end{array}\right]NJ\right)
\nonumber \\&&
-\frac{1}{2}\left[\begin{array}{c}a \\ a^\#\end{array}\right]^\dagger
\left(N^\dagger J N JP+PJN^\dagger J N\right)
\left[\begin{array}{c}a \\ a^\#\end{array}\right].
\end{eqnarray*}
\end{lemma}
{\em Proof:}
The proof of these identities follows via  straightforward but tedious
calculations using (\ref{CCR2}). \hfill $\Box$

\begin{lemma}
\label{L5}
With the variable $z$ defined as in (\ref{z01}) and $L$ defined as in (\ref{L}), then
\[
[z,L] = \left[E \left[\begin{array}{c}a \\ a^\#\end{array}\right], \tilde N\left[\begin{array}{c}a \\ a^\#\end{array}\right]\right] =  EJ\Sigma\tilde N^T
\]
which is a constant vector. Here, 
\[
\Sigma = \left[\begin{array}{cc}0 & I\\I & 0\end{array}\right].
\]
\end{lemma}
{\em Proof:}
The proof of this result  follows via  straightforward but tedious
calculations using (\ref{CCR2}). \hfill $\Box$

\begin{lemma}
\label{L6}
With the variable $z$ defined as in (\ref{z01}), $H_1$ defined in (\ref{H1}) and $L$ defined as in (\ref{L}), then
\begin{eqnarray*}
-i[z,H_1]+\mathcal{L}(z) &=&  E\left(-iJM-\frac{1}{2} JN^\dagger JN \right)
\left[\begin{array}{c}a \\ a^\#\end{array}\right]\\
&=&  E A \left[\begin{array}{c}a \\ a^\#\end{array}\right]
\end{eqnarray*}
where 
\begin{equation}
\label{A}
A = -iJM-\frac{1}{2} JN^\dagger JN. 
\end{equation}
Furthermore, 
\[
i[z,V] = 2i E JP\left[\begin{array}{c}a \\ a^\#\end{array}\right].
\]
\end{lemma}
{\em Proof:}
The proofs of these equations  follows via  straightforward but tedious
calculations using (\ref{CCR2}). \hfill $\Box$

We will  show that a sufficient condition for robust mean square
stability   when $H_2 \in \mathcal{W}_2$ is  the existence of a
constant $\theta \geq 0$, such that the following conditions are
satisfied:
\begin{enumerate}[(i)]
\item
The matrix $A$ defined in (\ref{A}) is Hurwitz.
\item
The transfer function 
\begin{equation}
\label{Gs}
G(s) =  -2i E \left(sI-A\right)^{-1} JE^\dagger
\end{equation}
satisfies the strict positive real (SPR) condition
\begin{eqnarray}
\label{SPR}
\gamma I - (1+\theta i \omega)G(i\omega) 
- (1-\theta i \omega)G(i\omega)^\dagger > 0
\end{eqnarray}
for all $ \omega \in [-\infty,\infty]$. 
\end{enumerate}
This leads to the following theorem.

\begin{theorem}
\label{T4}
Consider an uncertain open quantum system defined by $(S,L,H)$  such that
$H=H_1+H_2$ where $H_1$ is of the form (\ref{H1}), $L$ is of the
form (\ref{L}) and $H_2 \in \mathcal{W}_2$. Furthermore, assume that there exist a constant $\theta \geq 0$ 
such that the matrix $A$ defined in (\ref{A}) is Hurwitz and the frequency domain condition  (\ref{SPR})
is satisfied. Then the
uncertain quantum system is robustly mean square stable. 
\end{theorem}

\noindent
{\em Proof of Theorem \ref{T4}.}
If the conditions of the theorem are satisfied, then the transfer function
$\frac{\gamma}{2}I - (1+\theta s)G(s)$ is strictly positive real. However, this transfer function has a state space realization
\[
\frac{\gamma}{2}I - (1+\theta s)G(s) \sim 
\left[\begin{array}{c|c} A & B\\
\hline \\
-C - \theta C A & \frac{\gamma}{2}I -\theta C B 
\end{array}\right]
\]
where $A$ is defined as in (\ref{A}),  
\begin{equation}
\label{B}
 B = -2iJE^\dagger
\end{equation}
 and 
\begin{equation}
\label{C}
C =  E. 
\end{equation}
It now follows using the strict positive real lemma that the linear matrix inequality
\begin{equation}
\label{SPRLMI}
\left[\begin{array}{cc} 
PA + A^\dagger P & P B + C^\dagger + \theta A^\dagger C^\dagger \\
B^\dagger P + C+\theta C A & -\gamma I + \theta(CB+B^\dagger C^\dagger)\end{array} \right] < 0
\end{equation}
will have a solution $P > 0$ of the form (\ref{Pform}); e.g., see
\cite{KHA02}.  This matrix $P$ defines a corresponding operator $V \in \mathcal{P}_1$ as in (\ref{quadV}). Furthermore, it is straightforward to verify that $CB+B^\dagger C^\dagger = 0$. Hence, using Schur complements, it follows from (\ref{SPRLMI}) that
\begin{eqnarray}
\label{SPRQMI}
\lefteqn{PA + A^\dagger P}\nonumber \\
&&+ \frac{1}{\gamma}\left( P B + C^\dagger + \theta A^\dagger C^\dagger\right)
\left(B^\dagger P + C+\theta C A\right) < 0.\nonumber \\
\end{eqnarray}

Now using Lemma \ref{L6} we have
\begin{eqnarray*}
\lefteqn{i[z,V-\theta H_1]+\theta\mathcal{L}(z)+z}\\
 &=& i[z,V] +\theta\left(-i[z,H_1]+\mathcal{L}(z) \right)+z\\
&=& \left(2i EJP +\theta   E A  + E \right)\left[\begin{array}{c}a \\ a^\#\end{array}\right]\\
&=& E\left(2i JP +\theta  A  + I \right)\left[\begin{array}{c}a \\ a^\#\end{array}\right].
\end{eqnarray*}
Hence using Lemma \ref{L2}, we obtain
\begin{eqnarray}
\label{lyap_ineq3}
\lefteqn{-i[V,H_1]+ \mathcal{L}(V)} && \nonumber \\
&&+\frac{1}{\gamma}\left(i[z,V-\theta
  H_1]+\theta\mathcal{L}(z)+z\right)^\dagger \nonumber \\
&&\times \left(i[z,V-\theta H_1]+\theta\mathcal{L}(z)+z\right)
\nonumber \\ 
&&+ \frac{4\theta}{\gamma}[z,L]^\dagger [z,L]\nonumber \\
&=& \left[\begin{array}{c}a \\ a^\#\end{array}\right]^\dagger\tilde M \left[\begin{array}{c}a \\
a^\#\end{array}\right]\nonumber \\
&&+\tr\left(PJN^\dagger\left[\begin{array}{cc}I & 0 \\ 0 & 0 \end{array}\right]NJ\right)\nonumber \\
&&+ \frac{4\theta}{\gamma}\tilde N^\# \Sigma J E^\dagger EJ\Sigma\tilde N^T
\end{eqnarray}
where 
\begin{eqnarray*}
\lefteqn{\tilde M=}\\
&&PA + A^\dagger P+ \\
&&\frac{1}{\gamma}\left(2i JP +\theta A  + I \right)^\dagger E^\dagger 
 E\left(2i JP +\theta A  + I \right)\\
&=&PA + A^\dagger P\nonumber \\
&&+ \frac{1}{\gamma}\left( P B + C^\dagger + \theta A^\dagger C^\dagger\right)
\left(B^\dagger P + C+\theta C A\right),
\end{eqnarray*}
$A$ is defined in (\ref{A}), $B$ is defined in (\ref{B}) and $C$ is
defined in (\ref{C}). 
From this, it follows using  Lemma \ref{L5}, Lemma \ref{LA}, (\ref{SPRQMI}), and a similar argument to the proof of Theorem \ref{T2}
that 
\begin{eqnarray*}
\mathcal{G}(V+\theta H_2) \leq 
\left[\begin{array}{c}a \\ a^\#\end{array}\right]^\dagger\tilde M \left[\begin{array}{c}a \\
a^\#\end{array}\right] + \lambda
\end{eqnarray*}
where 
\begin{eqnarray*}
 \lambda &=& \tr\left(PJN^\dagger\left[\begin{array}{cc}I & 0 \\ 0 & 0 \end{array}\right]NJ\right)\\
&&+ \frac{4\theta}{\gamma}\tilde N^\# \Sigma J E^\dagger EJ\Sigma\tilde N^T \\
&\geq& 0.
\end{eqnarray*}
It follows from (\ref{SPRQMI}) that $\tilde M < 0$. 
Hence using (\ref{Delta_bound}), it follows that there exists a constant $c > 0$ such that the condition 
\[
\mathcal{G}(V+\theta H_2) + c\left(V+\theta H_2\right) \leq \lambda
\]
 is satisfied. Therefore, it follows from  Lemma \ref{L00}, Lemma \ref{LA}, (\ref{Delta_bound}) and  $P > 0$ that 
\begin{eqnarray}
\label{ms_stable1}
\lefteqn{\left< \left[\begin{array}{c}a(t) \\ a^\#(t)\end{array}\right]^\dagger \left[\begin{array}{c}a(t) \\ a^\#(t)\end{array}\right] \right> \leq}\nonumber \\
&&  e^{-ct}\left< \left[\begin{array}{c}a(0) \\ a^\#(0)\end{array}\right]^\dagger \left[\begin{array}{c}a(0) \\ a^\#(0)\end{array}\right] \right>\frac{\lambda_{max}[P+\frac{4\theta}{\gamma} E^\dagger E]}{\lambda_{min}[P]}\nonumber \\
&&+ \frac{\lambda}{c\lambda_{min}[P]}~~\forall t \geq 0.
\end{eqnarray} 
 Hence, the condition (\ref{ms_stable0}) is satisfied with $c_1 = \frac{\lambda_{max}[P+\frac{4\theta}{\gamma} E^\dagger E]}{\lambda_{min}[P]} > 0$, $c_2 = c > 0$ and $c_3 = \frac{\lambda}{c\lambda_{min}[P]} \geq 0$. 
\hfill $\Box$

\begin{observation}
\label{O1}
A useful special case of the above result occurs when the QSDEs describing the nominal open quantum linear system depend only on annihilation operators and not on the creation operators; e.g., see \cite{MaP3,MaP4}. This case corresponds to the case of $M_2 = 0$ and $N_2 =0$. Also, we assume that $E_2 = 0$. In this case, we calculate the matrix $A$ in (\ref{A}) to be
\[
A = \left[\begin{array}{cc}A_1 & 0\\0 & A_1^\#\end{array}\right]
\]
where $A_1 = -iM_1-\frac{1}{2}N_1^\dagger N_1$. Also, we calculate the transfer function matrix $G(s)$ in (\ref{Gs}) to be
\[
G(s) = -2\left[\begin{array}{cc} G_1(s) & 0\\0 & -G_1(s^*)^\#\end{array}\right]
\]
where $G_1(s) = iE_1\left(sI-A_1\right)^{-1}E_1^\dagger$. 

We now consider the case in which $G_1(s)$ is a SISO transfer
function. In this case, the condition that the matrix $A$ in (\ref{A})
is Hurwitz reduces to the condition that the matrix 
\begin{equation}
\label{A1}
A_1 = -iM_1-\frac{1}{2}N_1^\dagger N_1
\end{equation}
is Hurwitz. Also, the 
the SPR condition  (\ref{SPR}) reduces to the following conditions:
\begin{eqnarray}
\label{SPR2a}
\frac{\gamma}{4} +\mathcal{R}e[G_1(i\omega)] -\theta \omega\mathcal{I}m[G_1(i\omega)]  &>& 0;\\
\label{SPR2b}
\frac{\gamma}{4} -\mathcal{R}e[G_1(i\omega)] + \theta \omega \mathcal{I}m[G_1(i\omega)]  &>& 0
\end{eqnarray}
for all $ \omega \in [-\infty,\infty]$. The conditions (\ref{SPR2a}),
(\ref{SPR2b}) can be tested graphically producing a plot of $\omega
\mathcal{R}e[G_1(i\omega)]$ versus $\mathcal{I}m[G_1(i\omega)]$ with
$\omega\in [-\infty,\infty]$ as a parameter. Such a parametric plot
is referred to as the  Popov plot; e.g., see \cite{KHA02}. Then, the conditions (\ref{SPR2a}),
(\ref{SPR2b}) will be satisfied if and only if the Popov plot lies between
two straight lines of slope $\frac{1}{\theta}$ and with $x$-axis intercepts
$\pm \frac{\gamma}{4}$; see Figure \ref{F2}. 

\begin{figure}[htbp]
\begin{center}
\psfrag{Im}{$\omega\mathcal{I}m[G_1(i\omega)]$}
\psfrag{Re}{$ \mathcal{R}e[G_1(i\omega)]$}
\psfrag{slope}{slope $= \frac{1}{\theta}$}
\psfrag{g4}{$\frac{\gamma}{4}$}
\psfrag{mg4}{$-\frac{\gamma}{4}$}
\psfrag{ar}{allowable region}
\includegraphics[width=6cm]{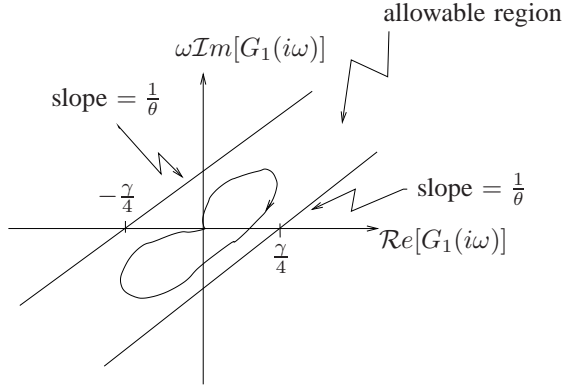}
\end{center}
\caption{Allowable region for the Popov plot.}
\label{F2}
\end{figure}
\end{observation}

\section{Illustrative Example}
\label{sec:example}
\begin{figure}[htbp]
\begin{center}
\psfrag{Im}{\tiny $\omega\mathcal{I}m[G_1(i\omega)]$}
\psfrag{Re}{\tiny $ \mathcal{R}e[G_1(i\omega)]$}
\includegraphics[width=8cm]{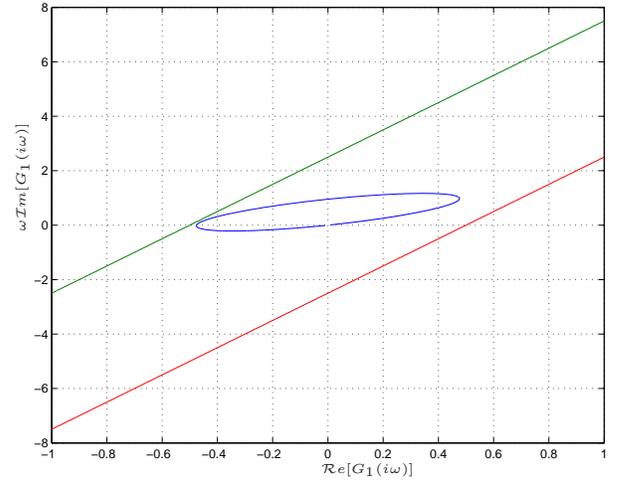}
\end{center}
\caption{Popov plot for the optical parametric amplifier system.}
\label{F3}
\end{figure}

In this section, we  consider an example of an open quantum system with 
\[
S=I,~H = \frac{1}{2}i\left(\left(a^*\right)^2-a^2\right),~L
= \sqrt{\kappa}a,
\]
which corresponds  an optical
parametric amplifier; see \cite{GZ00}. This is the same example which was considered in \cite{PUJ2}. In order apply the theory of this paper to  this example, we let
\[
H_2 = \frac{1}{2}[a^*~~a]\left[\begin{array}{cc}1 & i\\-i & 1\end{array}\right]\left[\begin{array}{c}a \\ a^* \end{array}\right] \geq 0
\]
and
\[
H_1 = \frac{1}{2}[a^*~~a]\left[\begin{array}{cc}-1 & 0\\0 & -1\end{array}\right]\left[\begin{array}{c}a \\ a^* \end{array}\right] 
\]
so that $H_1+H_2 = H$. 
This defines a linear quantum
system of the form considered in Theorem \ref{T4} with $M_1 = -1$, $M_2 =
0$, $N_1 = \sqrt{\kappa}$, $N_2 = 0$, $E_1=1$, $E_2 = 0$. Also,
$\Delta =\left[\begin{array}{cc}1 & i\\-i & 1\end{array}\right] \geq
0$, which  satisfies $\Delta \leq 2 I$. Hence, we can choose $\gamma =
2$ to ensure that condition (\ref{Delta_bound}) is satisfied and
  therefore $H_2   \in \mathcal{W}_2$. Also, note that this system is
  a system of the form considered in Observation \ref{O1} with $A_1 =
  i-\frac{\kappa}{2}$, which is Hurwitz for all $\kappa > 0$, and
  $G_1(s) = \frac{i}{s-i +\frac{\kappa}{2}}$. We then choose $\kappa =
  2.1$ and construct the  Popov plot corresponding to the
  transfer function $G_1(s)$ as
  discussed in Observation \ref{O1}. For a value of $\theta = 0.2$,
  this plot, along with the corresponding allowable region, is shown in
  Figure \ref{F3}. From this figure it can be seen that the 
  Popov plot lies within the allowable region and hence, it follows
  from Theorem \ref{T4} and Observation \ref{O1} that this system will be mean
square stable for $\kappa = 2.1$. However, the method of \cite{PUJ2}
could only prove that this system was stable for $\kappa > 4$. Hence
for this example, the proposed method provides a considerable
improvement over the method of  \cite{PUJ2}.

\section{Conclusions}
\label{sec:conclusions}
In this paper, we have considered the problem of robust stability for
uncertain  linear quantum systems with uncertain quadratic
perturbations to the system Hamiltonian. The stability condition which
is obtained is a quantum version of the classical Popov stability
criterion. This frequency domain condition is less conservative than a
previous stability result obtained which takes the form of a quantum
version of the classical small gain theorem. 

\end{document}